\documentclass[a4paper,11pt]{article}
\usepackage[dvipsnames]{xcolor}
\usepackage{pos}

\renewcommand{\logo}{\relax}
\graphicspath{ {figures/} }

\newcommand{\qhat}{\hat{q}}
\newcommand{\Msq}[1]{|\overline{\mathcal{M}}_{#1}|^2}

\title{Transport properties of the strongly interacting quark-gluon plasma}

\author*[a]{Ilia Grishmanovskii}
\author[b]{Taesoo Song}
\author[a,c]{Olga Soloveva}
\author[a,c]{Carsten Greiner}
\author[a,b,c]{Elena Bratkovskaya}

\affiliation[a]{Institut für Theoretische Physik, Johann Wolfgang Goethe-Universität,\\ Max-von-Laue-Straße 1, Frankfurt am Main, Germany}
\affiliation[b]{GSI Helmholtzzentrum für Schwerionenforschung GmbH,\\ Planckstraße 1, Darmstadt, Germany}
\affiliation[c]{Helmholtz Research Academy Hesse for FAIR (HFHF), GSI Helmholtz Center for Heavy Ion Physics,\\ Ruth-Moufang-Straße 1, Frankfurt am Main, Germany}

\emailAdd{grishm@itp.uni-frankfurt.de}
\emailAdd{t.song@gsi.de}
\emailAdd{soloveva@itp.uni-frankfurt.de}
\emailAdd{e.bratkovskaya@gsi.de}

\abstract{
We investigate the transport properties of the strongly interacting quark-gluon plasma (sQGP) by comparing the role of elastic and inelastic (radiative) processes in the sQGP medium within the effective dynamical quasi-particle model (DQPM), constructed for the description of non-perturbative quantum chromodynamic (QCD) phenomena of the sQGP in line with the lattice QCD (lQCD) equation of state. First, we present the results for the energy and temperature dependencies of the total radiative cross sections and compare them to the corresponding elastic cross sections. Second, we perform a calculation of the interaction rate and relaxation time of radiative versus elastic scatterings. Finally, we obtain the jet transport coefficient $\qhat$ and investigate its dependence on the choice of the strong coupling in thermal, jet parton and radiative vertices.
}

\FullConference{42nd International Conference on High Energy Physics (ICHEP2024)\\
18-24 July 2024\\
Prague, Czech Republic\\}

\begin{document}
\maketitle

Ultra-relativistic heavy-ion collisions performed at the Super Proton Synchrotron (SPS), the Relativistic Heavy-Ion Collider (RHIC) and the Large Hadron Collider (LHC) at CERN provide an access to the hot and dense phase of matter, the quark-gluon plasma (QGP). An understanding of the properties of the QGP is one of the main goals of current research in heavy-ion physics. 

A possible way to obtain information about the degrees of freedom of the QGP and its properties is the use of effective QCD models such as quasiparticle models (QPMs). Although QPMs rely on assumptions and external parameters, they focus on the most relevant degrees of freedom and interactions, making it possible to investigate the evolution of the QGP in dynamics.

The goal of this study is to investigate the transport properties of the QGP medium within the framework of the effective dynamical quasiparticle model (DQPM). The DQPM provides an effective approach to describe the QGP in terms of strongly interacting quarks and gluons, whose properties are fitted to match lattice QCD calculations in thermal equilibrium and at zero quark chemical potential. In the DQPM, quasiparticles are characterized by dressed propagators with complex self-energies, where the real part of the self-energies represents dynamically generated thermal masses, while the imaginary part carries information about the partons reaction rates. More details about the DQPM can be found in Refs. \cite{Peshier:2005pp, Grishmanovskii:2023gog}

\section{Partonic cross sections}

We start with investigating partonic interactions in the QGP medium. The partonic interaction cross sections in the DQPM are evaluated based on LO and NLO scattering diagrams with the use of the DQPM dressed propagators and couplings. In general, the DQPM cross sections depend on the temperature ($T$), baryon chemical potential ($\mu_B$), the invariant energy of the colliding partons ($\sqrt{s}$), and the masses of the participating partons.

\begin{figure}[h]
    \centering
    \includegraphics[width=0.49\textwidth]{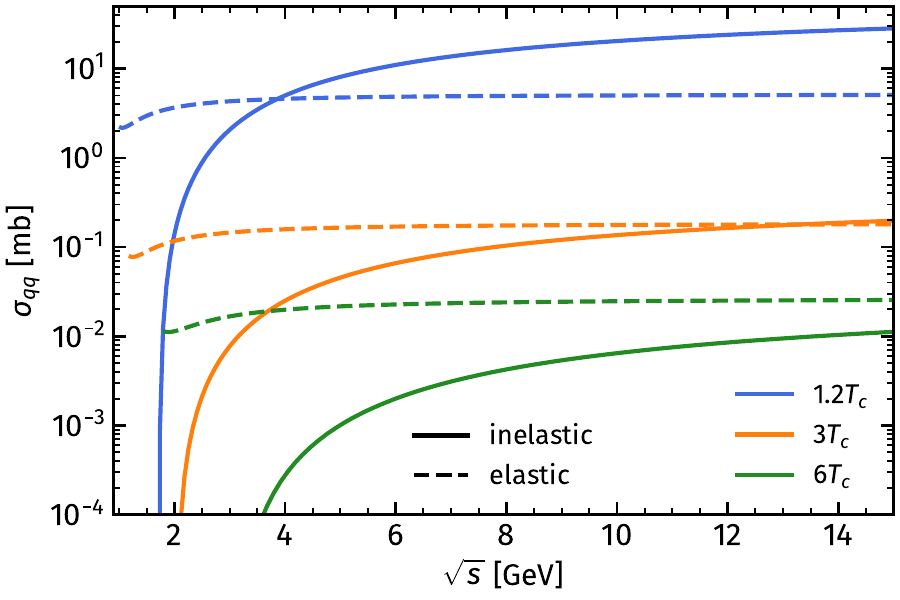}
    \includegraphics[width=0.49\textwidth]{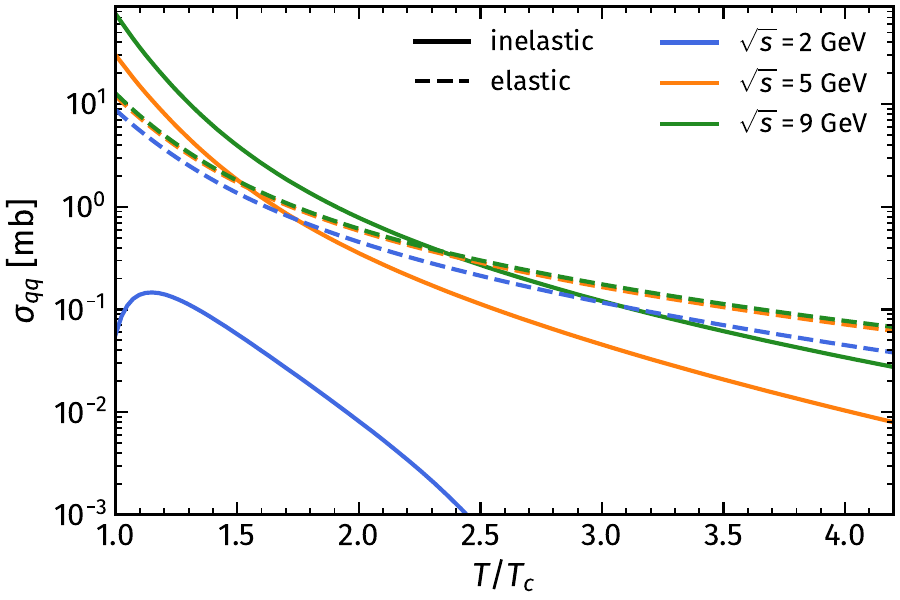}
    \caption{
        DQPM total elastic $qq \to qq$ and inelastic $qq \to qqg$ cross sections as functions of the collision energy $\sqrt{s}$ at different temperatures $T$ (left) and as functions of $T$ at different $\sqrt{s}$ (right).
    }
    \label{fig:XSi-S}
\end{figure}

In Fig. \ref{fig:XSi-S} we present the DQPM total elastic $qq \to qq$ and inelastic $qq \to qqg$ cross sections as functions of $\sqrt{s}$ at different $T$ and as functions of $T$ at different $\sqrt{s}$. Both processes are calculated for on-shell partons with initial parton masses taken as the pole masses of the spectral functions at a given temperature, as defined by the DQPM.

As seen from the figures, both elastic and inelastic cross sections increase monotonically with energy and decrease with temperature, but the shapes of these dependencies are different. The inelastic cross sections are generally suppressed compared to the elastic ones, especially for low energies and high temperatures, but become significant in the low-temperature regime at high energies. The reason for such a behavior of the DQPM cross section comes from the form of the scattering amplitudes, which are proportional to the coupling squared ($\Msq{2 \to 2} \propto \alpha_s^2$) for the elastic reactions and are proportional to the coupling cubed ($\Msq{2 \to 3} \propto \alpha_s^3$) for the inelastic reactions. Since $\alpha_s$ in the DQPM strongly depends on temperature and grows above 1 in the vicinity of the critical temperature (see \cite{Grishmanovskii:2023gog} for details), it leads to a more rapid increase of the inelastic amplitudes at low temperatures. 

\section{Partonic interaction rates}

In addition to the partonic cross section, another quantity of interest in the context of partonic interactions is the partonic interaction rate, which describes the frequency at which partons interact with each other within the QGP medium. 

\begin{figure}[h]
    \centering
    \includegraphics[width=0.49\textwidth]{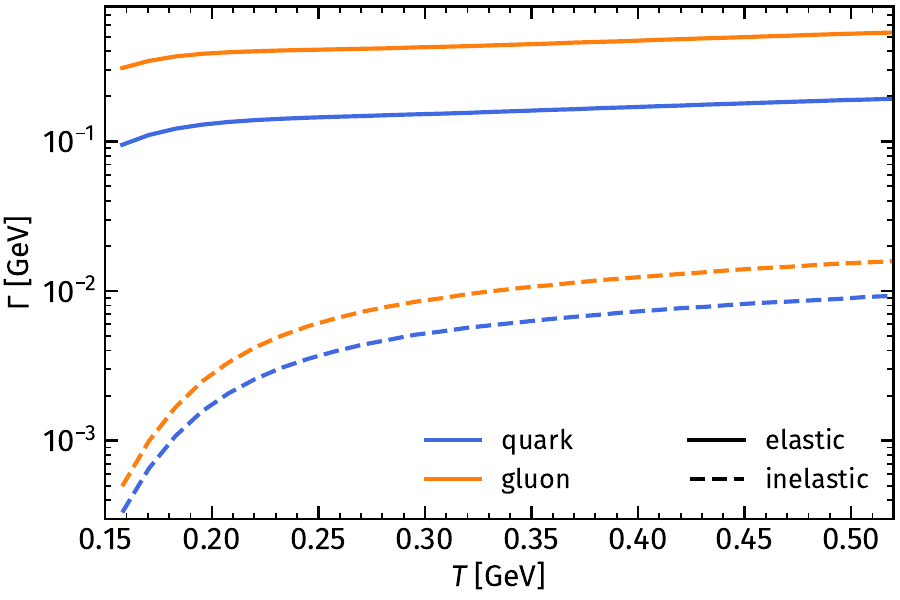}
    \includegraphics[width=0.49\textwidth]{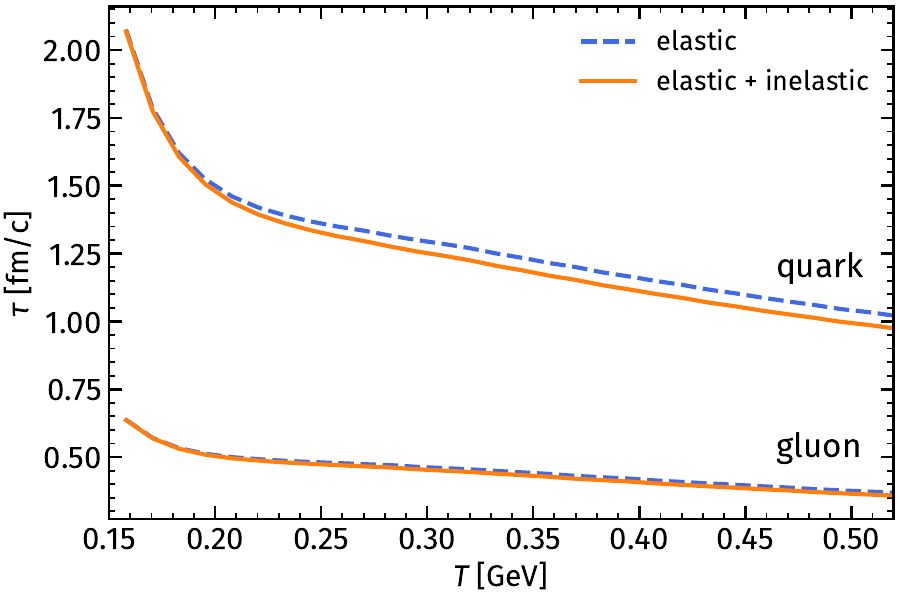}
    \caption{
        Left: Interaction rates $\Gamma$ for a light quark and a gluon as functions of the temperature $T$ at $\mu_B = 0$ for elastic and inelastic processes.
        Right: Relaxation times $\tau$ of a light quark and gluon as functions of the temperature $T$ at $\mu_B = 0$, evaluated by the average partonic interaction rate for the elastic contribution only and for the sum of elastic and inelastic contributions.
    }
    \label{fig:IR_RT}
\end{figure}

The left-hand side of Fig. \ref{fig:IR_RT} shows the interaction rate $\Gamma$ for a light quark and a gluon as a function of the temperature $T$ at $\mu_B = 0$ for elastic and inelastic processes. As seen from the figure, the inelastic interaction rates are strongly suppressed at all temperatures compared to the elastic rates. This suppression can be explained if one considers the distribution of scattered quasiparticles inside the QGP medium. Due to the distribution function, which suppresses collisions at large $\sqrt{s}$, only the low-$\sqrt{s}$ region of elastic and inelastic cross sections contributes dominantly to the interaction rates. Since in this region the inelastic reactions are essentially suppressed, it leads to the suppression of the inelastic interaction rates, indicating that the contribution of inelastic processes -- with the emission of massive thermal gluons -- is negligible for the thermal properties of the sQGP within the quasiparticle picture.

The interaction rate can be directly related to the relaxation time $\tau = \hbar c/\Gamma$, which is defined as the time it takes for a system to return to equilibrium after a change or disturbance. The right-hand site of Fig. \ref{fig:IR_RT} depicts the relaxation time for a light quark and gluon as a function of $T$ for elastic contributions only and for the sum of elastic and inelastic contributions. It is seen that accounting for inelastic processes just slightly shortens the relaxation time of thermal sQGP partons.

\section{Jet transport coefficients}

We proceed with studying the interaction of a fast jet parton with the QGP medium. The propagation of the jet particle can be described in terms of the jet transport coefficients, which can be calculated within the DQPM with the use of kinetic theory \cite{Moore:2004tg} by employing the Boltzmann transport equation. In this study we are interested in the $\qhat$ coefficient \cite{Grishmanovskii:2022tpb, Grishmanovskii:2024gag}, which characterizes the transverse momentum broadening per unit length of a high-energy parton traversing the medium.

\begin{figure}[!ht]
    \renewcommand{\arraystretch}{1.3}
    \centering
    \begin{minipage}[t]{0.3\textwidth}
        \includegraphics[width=0.9\textwidth]{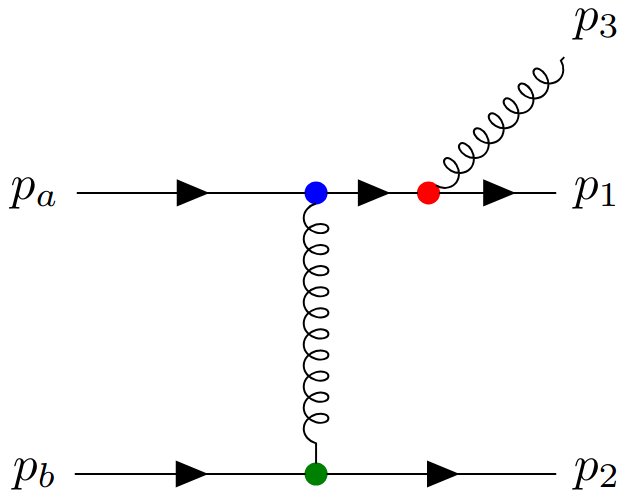}
    \end{minipage}
    \hfill
    \begin{minipage}[t]{0.6\textwidth}
        \vspace*{-3.3cm}
        \scalebox{0.8}{
        \begin{tabular}{|l|c|c|c|}
        \hline
        & \multicolumn{3}{c|}{\textbf{Vertex}}
        \\ \hline
        \textbf{Model} & \textcolor{Green}{$\bullet$} medium parton\; & \textcolor{blue}{$\bullet$} jet parton\; & \textcolor{red}{$\bullet$} emitted gluon\;
        \\ \hline \hline
        Scenario 0 & \multicolumn{3}{c|}{$g^{\mathrm{DQPM}}(T)$}
        \\ \hline
        Scenario I & \multicolumn{3}{c|}{$g = \sqrt{4 \pi \times 0.3}$}
        \\ \hline
        Scenario II & $g^{\mathrm{DQPM}}(T)$ & $g(Q^2)$ \cite{Zakharov:2020whb} & $g(k_t^2)$ \cite{Zakharov:2020whb} 
        \\ \hline
        Scenario III & $g^{\mathrm{DQPM}}(T)$ & $g(E)$ \cite{Liu:2023rfi} & $g(E)$ \cite{Liu:2023rfi} 
        \\ \hline
        Scenario IV & $g^{\mathrm{DQPM}}(T)$ & $g(ET)$ \cite{Karmakar:2023ity} & $g(Q^2)$ \cite{Karmakar:2023ity} 
        \\ \hline
        \end{tabular}
        }
    \end{minipage}
    \caption{
        Left: An example of a Feynman diagram for a $qq \to qqg$ reaction. Colored dots indicate different vertices.
        Right: Scenarios for the effective coupling.
    }
    \label{tbl:scenarios_alphas}
\end{figure}

Since a jet parton is not part of the QGP medium, it is unclear whether the strong couplings, associated with the jet and the emitted gluon, should remain thermal, considering that the gluon can originate from a non-thermal jet parton. Hence, in this section we examine five scenarios, motivated by Refs. \cite{Zakharov:2020whb, Liu:2023rfi, Karmakar:2023ity}, involving different strong couplings for every possible vertex: one associated with the thermal parton, one associated with the jet parton, and one associated with the emitted gluon. An overview of the presented scenarios is shown in Fig. \ref{tbl:scenarios_alphas}. The detailed description can be found in Ref. \cite{Grishmanovskii:2024gag}. We note that for the elastic reactions the thermal and jet vertices are the same as in the corresponding Feynman diagrams for the inelastic case.

\begin{figure}[ht!]
    \centering
    \includegraphics[width=0.5\textwidth]{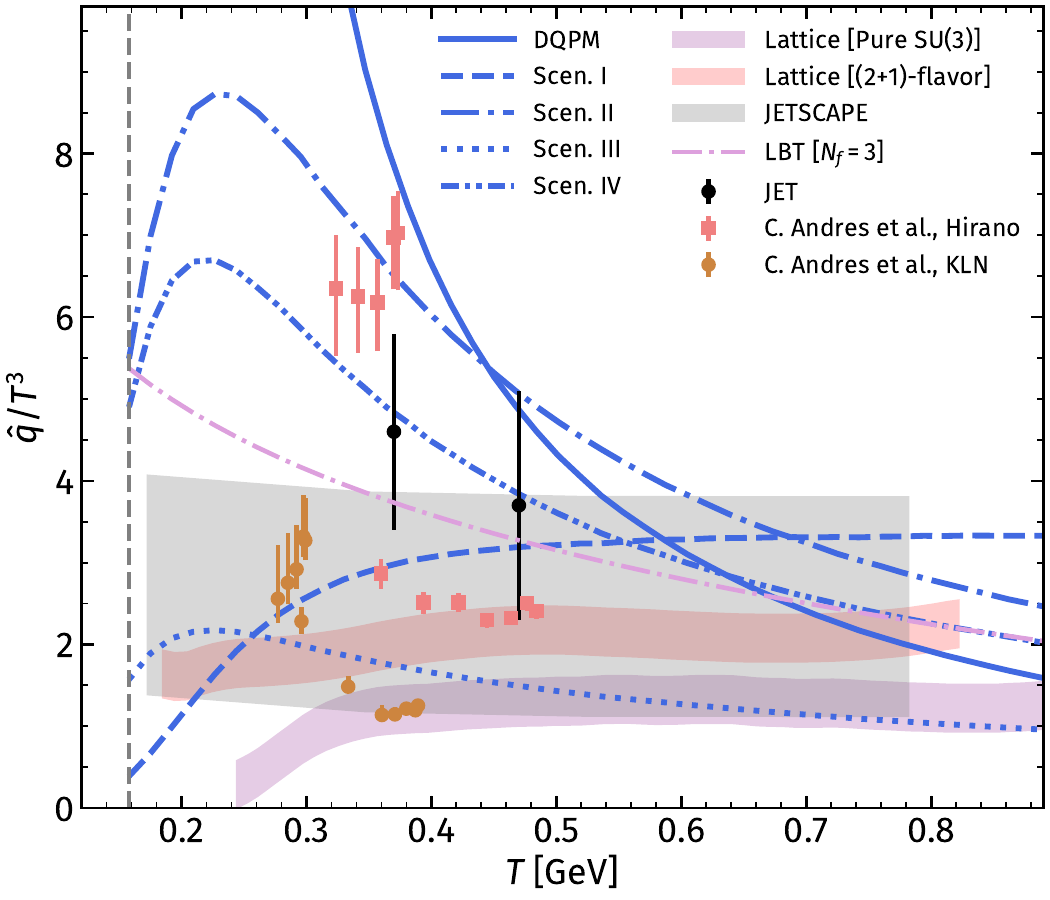}
    \caption{
        Temperature dependence of the scaled jet transport coefficient $\qhat/T^3$. Blue lines depict the DQPM results for a quark jet with mass $M_j = 0.01$ GeV and energy $E = 10$ GeV for different scenarios (see text). The pink dash-dotted line demonstrates the LBT results for $N_f = 3$ and $p = 10$ GeV$/c$ \cite{He:2015pra}. The red (upper) and purple (lower) areas show the lQCD estimates \cite{Kumar:2020wvb} for pure SU(3) gauge theory and (2+1) flavor QCD, respectively, in the limit of an infinitely fast jet parton. The gray area corresponds to the results from the JETSCAPE Collaboration ($p = 100$ GeV$/c$) \cite{JETSCAPE:2021ehl}. The black dots represent the phenomenological extraction by the JET Collaboration presented for $p = 10$ GeV$/c$ \cite{Burke:2013yra}. The orange and red symbols display results from Ref. \cite{Andres:2016iys}. The vertical gray dashed line indicates the critical temperature $T_c = 0.158$ GeV.
    }
    \label{fig:QHAT}
\end{figure}

The results for the temperature dependence of the elastic + inelastic $\qhat/T^3$ coefficient for scenarios 0-IV (for the momentum of a jet parton of $p = 10$ GeV$/c$) are presented in Fig. \ref{fig:QHAT} with the blue lines and compared to various models (see legend). A large discrepancy between different scenarios indicates that the behavior of $\qhat$ is primarily dominated by the choice of the strong coupling $g$, especially at low temperatures. While Scenario I, with a fixed $\alpha_s = 0.3$, shows a decrease of $\qhat$ at low $T$, the DQPM (Scenario 0) demonstrates the strongest rise due to the thermal $g(T)$ for all vertices. Other Scenarios II-IV include the momentum (energy) dependence of the strong coupling in jet and radiative vertices, thereby modifying the $T$ as well as $E$ dependencies of $\qhat$ (see Ref. \cite{Grishmanovskii:2024gag} for details) and suppressing the rapid rise at low temperatures, observed in Scenario 0.

One can also see from Fig. \ref{fig:QHAT} that the spread of different model results for $\qhat$ is very large, and this discrepancy generally depends on both the model assumptions and the method used for extracting $\qhat$ from the heavy-ion data, particularly from flow coefficients $v_n$ and the ratio $R_{AA}$.

\section{Summary}

We have studied the transport properties of the strongly interacting QGP medium within the DQPM model, which describes the QGP in terms of strongly interacting massive off-shell quarks and gluons, whose properties are fitted to reproduce lQCD EOS in equilibrium. The comparison of the elastic and inelastic cross sections shows that while elastic cross sections dominate at low energies and high temperatures, radiative cross sections increase at low temperatures and high energies, becoming comparable or even dominant. This behavior originates from the behavior of the DQPM strong coupling. We have also found that the interaction rate and relaxation time in the sQGP are primarily governed by elastic scattering, suggesting that inelastic processes with massive gluon emission are suppressed in the nonperturbative thermalized sQGP medium. Finally, we have studied the interaction of leading jet partons with the QGP medium by calculating the jet transport coefficient $\qhat$. The calculation of $\qhat$ within different scenarios for the strong couplings has revealed a high sensitivity of the transport coefficients on the choice of energy, momentum, or temperature dependencies of the strong couplings. The "default" DQPM with the thermal couplings produces the highest values of the transport coefficients.

\acknowledgments

We acknowledge support by the Deutsche Forschungsgemeinschaft (DFG, German Research Foundation) through the grant CRC-TR 211 "Strong-interaction matter under extreme conditions" -- project number 315477589 -- TRR 211. 


\bibliographystyle{JHEP}
\bibliography{ICHEP24}

\end{document}